\documentclass[12pt]{article}
\usepackage{epsfig,amsfonts,amssymb}
\usepackage{hyperref}
\usepackage{cite}
\usepackage{cite}

\def\ZZZ{{\hbox{ Z\kern-1.6mm Z}}}
\def\RRR{{\hbox{ R\kern-2.4mm R}}}
\def\CCC{{\hbox{ C\kern-2.5mm C}}}
\def\zzz{{\hbox{z\kern-1mm z}}}

\def\RRR{{\cal R}}

\newcommand{\qeq}{{\hbox{=\kern-2.3mm ? \kern.5mm }}}
\renewcommand{\qeq}{=}

\newcommand{\eps}{\epsilon}

\newcommand{\ve}{\varepsilon}

\newcommand{\KK}{{\cal K}}

\newcommand{\MM}{{\cal M}}

\newcommand{\OO}{{\cal O}}

\newcommand{\EE}{{\cal E}}
\newcommand{\LL}{{\cal L}}

\newcommand{\wt}{\widetilde}
\newcommand{\wh}{\widehat}

\newcommand{\RR}{{\cal R}}

\newcommand{\SSS}{{\cal S}}

\newcommand{\be}{\begin{equation}}
\newcommand{\ee}{\end{equation}}
\newcommand{\ben}{\begin{eqnarray}\displaystyle}
\newcommand{\een}{\end{eqnarray}}

\newcommand{\refb}[1]{(\ref{#1})}
\newcommand{\p}{\partial}
\newcommand{\sectiono}[1]{\section{#1}\setcounter{equation}{0}}

\def\one{{\hbox{ 1\kern-.8mm l}}}
\def\zero{{\hbox{ 0\kern-1.5mm 0}}}

\newcommand{\bea}[1]{\begin{eqnarray}\label{#1} }
\newcommand{\eea}{\end{eqnarray}}

\newcommand{\eqref}{\refb}

\usepackage{bm}
\usepackage[table]{xcolor}

\begin{document}

\vskip 12pt

\baselineskip 24pt

\begin{center}


{\Large \bf Self-dual forms: Action, Hamiltonian and Compactification}

\end{center}

\def\RR{H}

\vskip .6cm
\medskip

\vspace*{4.0ex}

\baselineskip=18pt

\begin{center}

{\large 
\rm Ashoke Sen}

\end{center}

\vspace*{4.0ex}

\centerline{ \it  Harish-Chandra Research Institute, HBNI,}
\centerline{\it Chhatnag Road, Jhusi,
Allahabad 211019, India}

\vspace*{1.0ex}
\centerline{\small E-mail: sen@hri.res.in}

\vspace*{5.0ex}

\baselineskip=18pt

\centerline{\bf Abstract} \bigskip

It has been shown that,
by adding an extra free field that decouples from the dynamics, 
one can construct actions for interacting 2n-form fields 
with self-dual
field strengths in 4n+2 dimensions. 
In this paper we analyze canonical formulation of these theories, and show that
the resulting Hamiltonian reduces to
the sum of two Hamiltonians with independent degrees of freedom. 
One of them is free and has no physical
consequence, while the other contains the physical degrees of freedom with the desired interactions.
For the special cases of chiral scalars in two dimensions and chiral two form fields in six dimensions,
we discuss compactification of these theories respectively on a circle and a two dimensional torus, and show that we 
recover the expected properties of these systems, including S-duality invariance in four
dimensions.



\vfill \eject

\baselineskip=18pt

\tableofcontents

\sectiono{Introduction and summary} \label{s1}

Usual formulation of the theory of self-dual $(2n+1)$-form field strength in $4n+2$
dimensions involves writing an action of a $2n$-form potential 
without imposing any self-duality constraint on the field strength,
and imposing the self-duality constraint {\it after deriving the equations of motion from this action}. Various
approaches have been suggested for avoiding this problem. 
These
formulations either break manifest Lorentz 
invariance\cite{Henneaux,9701008,0605038}, 
or have infinite number
of auxiliary fields\cite{Mcclain,Wotzasek,Martin,9603031,Faddeev,9609102,9607070,9610134,9610226}, 
or have a finite number of auxiliary fields with non-polynomial  
action\cite{9509052,9611100,9701037,9701149,9707044,
9806140,9812170}, or require going to one higher 
dimension\cite{9610234,9912086}. Other attempts in this direction can be
found in \cite{cast1,cast2}.

However many such theories appear in compactification of type IIB string theory.
For the latter we can
write down an action at the cost of adding an extra free field, but without having to break manifest
Lorentz invariance, or introduce infinite number of
auxiliary fields (at massless level), or have non-polynomial action with scalar fields in the denominator\cite{1508.05387}. 
Therefore we expect that similar actions can also be written
down for general  self-dual
$(2n+1)$-form field strength in $4n+2$ dimensions. Indeed such an action for ten dimensional 
type IIB supergravity was constructed in \cite{1511.08220}, and generalization of this to 
$(2n+1)$-form field strength in $(4n+2)$ dimensions for general $n$ is 
straightforward. 

The
goal of this paper will be to explore this formulation in more detail. One of the somewhat puzzling features
of the action described in \cite{1511.08220} is the existence of an extra set of free fields which decouple from
the interacting degrees of freedom and therefore have no physical relevance. This decoupling can be seen at
the level of equations of motion but not at the level of the action. By embedding this formulation in
superstring field theory one can also show that the decoupling continues to hold to all orders in perturbation
theory\cite{1508.05387}. 
Nevertheless one would like to see this decoupling more explicitly. In this paper we show that if we
carry out the canonical formulation of the theory described by the action given 
in \cite{1511.08220}, the
corresponding Hamiltonian becomes a sum of two Hamiltonians, one describing a free theory and the other
describing an interacting theory.
Furthermore the Dirac brackets between the set of degrees of freedom
involved in the two Hamiltonians vanish, so that in the quantum theory the two Hamiltonians commute. This
gives a clear demonstration of the
separation between the extra free degrees of freedom and the interacting degrees of freedom.

We also study compactification of these theories on tori. 
Since the coupling of the metric to these theories is non-standard -- not through the usual covariantization
procedure -- these compactifications avoid some of the apparent 
no go theorems\cite{0905.2720,1903.02825}.
We analyze two cases in detail, the compactification
of chiral boson in two dimensions on a circle and a chiral 2-form in six dimensions on $T^2$. 

We shall conclude this section by summarizing the results described in different sections 
of the paper. 
In \S\ref{s2} we give a brief review of the results
of \cite{1511.08220} for the construction of an action of $2n$-form fields with self-dual field strength
in $(4n+2)$ dimensions, at the cost of introducing
an extra free field that decouples from the rest of the
degrees of freedom. In \S\ref{s3} we analyze the canonical formulation of the theory. For simplicity
we treat the degrees of freedom associated with the $2n$-form fields
and $(2n+1)$-form field strengths in the canonical formalism
but the rest of the degrees of freedom, including the metric, are treated in the Lagrangian 
formalism. Therefore what we obtain at the
end can be identified as the Routhian instead of the Hamiltonian, although we shall continue to
refer to it as the Hamiltonian. Alternatively we can interpret our result as the Hamiltonian of the
$2n$-form fields  in the background of other fields.
The final result $\RR_++\RR_-$, with $\RR_\pm$ 
given in \refb{er+r-}, 
is a sum of two decoupled systems.
One of them is 
a quadratic function of the variable $\Pi_+$ and the other is a complicated function of the
variables $\Pi_-$ and the  rest of the degrees of freedom, including the metric, collectively denoted
by $\Psi$. The associated equations of motions are given in \refb{ehameq},
\refb{erouth}, with $\Pi_\pm$ satisfying
the Dirac bracket relations \refb{ediracb} and the constraints \refb{e3.17}. From this the decoupling
of $\Pi_+$ from the rest of the degrees of freedom is obvious. 

In \S\ref{s4} we study the special
case of $n=0$, corresponding to a chiral boson in two dimensions and its compactification on a 
circle of radius $R$. In this case the free part 
$\RR_+$ takes the usual form given in \refb{edegh2d}, 
while the interacting
part $\RR_-$ is shown  to be given by \refb{errmfin}. Treating the background metric as classical
we can quantize the degrees of freedom $\Pi_\pm$ by replacing the Dirac bracket by 
$-i$ times the commutator bracket. The spectrum of the resulting quantum Hamiltonian $\RR_-$,
given in
\refb{echiralH}, agrees with the expected spectrum of a chiral boson compactified on a circle of
radius $R$.

In \S\ref{s5} we analyze the case $n=1$, corresponding to the theory of 2-forms with self-dual
3-form
field strength in six dimensions. The action of the theory is taken to be the one given in
\refb{e5fin}, which can be shown to be equivalent to the action \refb{e5.1} 
together with a self-duality
constraint \refb{e5con}. Instead of considering full compactification of the theory, we 
consider its dimensional reduction on $T^2$ by throwing away the Kaluza-Klein modes. We also
carry out a consistent truncation in which we throw away the scalar fields arising out of the 
compactification, but keep the vector fields and their sources.  The resulting interacting 
Hamiltonian 
$\RR_-$ is given 
by \refb{es5hamfin}, with 
$\Pi_-$ satisfying the constraint \refb{econ6d} and Dirac brackets \refb{ediracb6d}. The 
Hamiltonian, 
as well as the constraints and the Dirac bracket relations, 
are manifestly invariant under the S-duality
transformation given in \refb{eduality}.

\sectiono{Review} \label{s2}

In this section we shall review the construction of \cite{1511.08220} for the action of a 
$2n$-form field, with self-dual $(2n+1)$-form field strength, in $4n+2$ dimensions.
The field content of the theory is a $2n$-form field $P$, a self-dual $(2n+1)$-form 
field $Q$
satisfying the algebraic relation
\be
*Q=Q,
\ee
and other fields that we shall collectively denote by $\Psi$. $*$ denotes 
{\it Hodge duality with respect to flat Minkowski metric $\eta_{\mu\nu}$}. 
The action takes the form
\be\label{eact}
S = {1\over 2} \int d P\wedge *dP -  \int d P\wedge Q  + \int \LL_I(Q, \Psi)\, ,
\ee
where $\LL_I$ is some Lagrangian density 
that depends on $Q$ and the other fields $\Psi$ but not on $P$. The equations of motion
take the form:
\be\label{esec}
d (*dP - Q) = 0, \quad dP -*dP + R(Q,\Psi) = 0, \quad {\delta\over \delta \Psi}\int \LL_I 
=0\, ,
\ee
where $R$ is an anti-self-dual $(2n+2)$-form defined so that under a variation of $Q$, 
\be \label{edefR}
\delta \int \LL_I = -{1\over 2} \int R\wedge \delta Q\, .
\ee

Throughout the paper all indices will be raised and
lowered by flat Minkowski metric.
For generic $(2n+1)$-forms $A$ and $B$,
our conventions for the wedge product and Hodge dual are as follows:
\be\label{e2.2}
\int A\wedge B = \int d^{4n+2} x \, \eps^{\mu_1\cdots \mu_{2n+1} \nu_1\cdots \nu_{2n+1}}
\, A_{\mu_1\cdots \mu_{2n+1}} \, B_{\nu_1\cdots \nu_{2n+1}}\, ,
\ee
\be \label{e2.3}
(*A)_{\mu_1\cdots \mu_{2n+1}} = \eta_{\mu_1\nu_1}\cdots \eta_{\mu_{2n+1}\nu_{2n+1}}\,
\eps^{\nu_1\cdots \nu_{2n+1} \rho_1\cdots \rho_{2n+1}} \, A_{\rho_1\cdots
\rho_{2n+1}} \, ,
\ee
where $\eps^{\mu_1\cdots \mu_{4n+2}}$ is totally anti-symmetric in all the indices and
$\eps^{01\cdots (4n+1)}=1$. 
In the above equations the contraction between a pair of anti-symmetric sets of indices,
{\it e.g.} $\{\mu_1,\cdots,\mu_{2n+1}\}$ and $\{\nu_1,\cdots,\nu_{2n+1}\}$ in \refb{e2.2}
and $\{\rho_1,\cdots,\rho_{2n+1}\}$ in \refb{e2.3},
runs only over
inequivalent combinations, {\it e.g.} over 
$\mu_1<\mu_2<\cdots <\mu_{2n+1}$ and $\nu_1<\nu_2<\cdots <\nu_{2n+1}$ in \refb{e2.2}
and over $\rho_1<\rho_2<\cdots <\rho_{2n+1}$ in \refb{e2.3}. We also use the convention:
\be
(dP)_{\mu_1\cdots \mu_{2n+1}} = \p_{\mu_1} P_{\mu_2\cdots \mu_{2n+1}} +
(-1)^P \times \hbox{cyclic permutations of} \, \, \mu_1,\cdots, \mu_{2n+1}\, .
\ee

Expressing the first equation in \refb{esec}
as $d(dP+*dP-Q)=0$, we see that $dP+*dP-Q$ describes
a self-dual $(2n+1)$ form with vanishing exterior derivative. This describes the
degrees of freedom of a free
$2n$ form field in $4n+2$ dimensions with self-dual field strength. 
From the sign of the first term in \refb{eact}
we see that this mode has wrong sign kinetic term. However since this is a free
field it decouples from the system and is harmless. 
This will be seen more explicitly in the canonical formalism in \S\ref{s3}.
On the other hand, adding the
exterior derivative of the second equation in \refb{esec} to the first equation
we get 
\be \label{eqeq}
d(Q -R)=0\, .
\ee 
In the free theory $R$ vanishes and 
the equation describes a self-dual field $Q$ with vanishing exterior derivative. This
gives a second free field with self-dual field strength $Q$. 
However for $R\ne 0$ this is an interacting field,
with degrees of freedom still given by that of a $2n$ form field with self-dual field
strength.

For simplicity in this paper we shall work with $\LL_I$ that depends on $Q$ but not derivatives
of $Q$. However the general argument showing the decoupling of one set of free field degrees
of freedom from the rest continue to hold even when we include in $\LL_I$ derivatives of $Q$.

Note that the decoupling between the free field and the interacting fields  happens
at the level of the equations of motion but not at the level of the action. Nevertheless
a similar mechanism in string field theory has been shown to maintain this 
decoupling to all orders in perturbation theory\cite{1508.05387}. 
One of the goals in this paper will be to verify this decoupling
using canonical formulation of the theory described in \refb{eact}. 
We shall see that in this case the Hamiltonian of the
system reduces to the sum of a free Hamiltonian and an interacting Hamiltonian with
independent degrees of freedom. The free Hamiltonian has no lower bound on the
energy, but it has no physical consequence since it remains decoupled from the
physical degrees of freedom that enter the interacting Hamiltonian.

We shall now describe how to relate the action \refb{eact} to the usual formulation of self-dual 
$p$-form theories. There instead of the pair of fields $P,Q$ we introduce a $2n$-form potential $C$
and the $(2n+1)$-form field strength
\be \label{e2.2rep}
F \equiv d C\, . \ee
We now consider an action of the form
\be \label{e4.0}
S=S_1+S_2(\Psi)\, ,
\ee
\be \label{e4.1}
S_1 \equiv -{1\over 2} \int  (F+Y) \wedge *_g  (F+Y) + \int F \wedge Y\, ,
\ee
where $Y$ is a $(2n+1)$-form constructed out of the other fields $\Psi$
and $S_2$ is a function of the other fields $\Psi$, including the metric $g_{\mu\nu}$.
$*_g$ denotes Hodge dual with respect to the metric $g_{\mu\nu}$:
\be 
(*_g A)_{\mu_1\cdots \mu_{2n+1}} = (-\det g)^{-1/2}\,
g_{\mu_1\nu_1}\cdots g_{\mu_{2n+1}\nu_{2n+1}}\,
\eps^{\nu_1\cdots \nu_{2n+1} \rho_1\cdots \rho_{2n+1}} \, A_{\rho_1\cdots
\rho_{2n+1}} \, .
\ee
In addition to the equations of motion derived from the action $S$, we impose the self duality constraint
\be \label{e4con}
*_g (F+Y) = (F+Y)\, .
\ee
If we only want to study the coupling of the field to gravity, 
we can set $Y=0$. However we have included it in the action
explicitly since $dY$ acts as a source for the self-dual form $F$.
Also many supergravity
theories naturally have such couplings.

It was shown in \cite{1511.08220} 
that the action \refb{e4.0}, together with the constraint \refb{e4con}, is equivalent to 
a new action of the type described in \refb{eact}:
\be\label{e4tot}
S'=S_1'+S_2(\Psi)\, ,
\ee
where $S_2$ is the same action as in \refb{e4.0} and
\ben \label{efin}
S'_1 &=& {1\over 2}  \int d P \wedge * d P
- \int d P \wedge Q  \\ && 
+ \int d^{4n+2}x \, \left[{1\over 16}  Q^T     \MM \,
 Q  
+ {1\over 2}   Q^{T}  \bigg\{{1\over 2} \MM \,
Y - (\zeta-\ve) \,
Y\bigg\} 
-{1\over 2} Y^T 
\zeta \,Y
+{1\over 4} Y^T  \MM\,  Y\right] \, .  \nonumber 
\een
Here $P$ is a $2n$-form field and $Q$ is a self-dual $(2n+1)$-form field as before. 
$\MM$, $\zeta$ and
$\ve$ are  defined as follows. If $g_{\mu\nu}$ is the metric, we define $\hat e$ to be its symmetric square root
and $\hat E$ to be the inverse of $\hat e$ via the equations:
\be
\hat e^T=\hat e, \quad \hat e\, \eta \, \hat e = g, \quad \hat E = \hat e^{-1} \, .
\ee
If we define
\be 
h_{\mu\nu}=g_{\mu\nu}-\eta_{\mu\nu},
\ee
then $\hat e$ may be constructed as a series expansion in $h$:
\be 
\hat e = (1 + h\eta)^{1/2} \eta= \left(\eta + {1\over 2} h - {1\over 8} h\eta h + \cdots\right)
 \, .
\ee
$\hat e\eta$ plays the role of the vielbein, but
since the local Lorentz symmetry has already been gauge fixed by choosing $\hat e$ to be
symmetric, we shall not distinguish between the tangent space indices and coordinate indices
and use the  Greek indices $\mu,\nu,\cdots$ for both. As already mentioned before, all indices
will be raised and lowered by $\eta$.
We now introduce indices 
$A,B$ that correspond to completely anti-symmetric set of $(2n+1)$ space-time indices 
$[\mu_1\cdots \mu_{2n+1}]$, $[\nu_1\cdots \nu_{2n+1}]$. 
Clearly $A,B$ take ${4n+2\choose 2n+1}$ possible values.
We shall choose the convention that the indices $\mu_1,\cdots ,\mu_{2n+1}$ of $A$ are always
arranged as $\mu_1<\mu_2<\cdots <\mu_{2n+1}$ and that the sum over $A$ means sum over
$\mu_1,\cdots ,\mu_{2n+1}$ subject to this restriction.\footnote{This convention is slightly different
from the one used in \cite{1511.08220} where we 
had summed over all $\mu_1,\cdots \mu_{2n+1}$ and then
divided the result by $(2n+1)!$. Due to this different convention we need to include explicit 
anti-symmetrization in the expressions for $e_{AB}$ and $E^{AB}$ in \refb{evdef}. Similar 
anti-symmetrization is not needed for $\zeta$ since $\eta$ has no off-diagonal components.
}
We now define
\ben \label{evdef}
&& \zeta^{AB} = \eta^{\mu_1 \nu_1}\cdots \eta^{\mu_{2n+1} \nu_{2n+1}}, \quad \zeta_{AB} = 
\eta_{\mu_1 \nu_1}\cdots \eta_{\mu_{2n+1} \nu_{2n+1}}\, , \nonumber \\ && 
e_{AB} = \hat e_{\mu_1 \nu_1} \cdots \hat e_{\mu_{2n+1} \nu_{2n+1}}
+ \hbox{sum over permutations $P$ of 
$\nu_1,\cdots, \nu_{2n+1}$ weighted by $(-1)^P$} \, , \nonumber \\
&&
E^{AB} 
= \hat E^{\mu_1 \nu_1} \cdots \hat E^{\mu_{2n+1} \nu_{2n+1}} + \hbox{sum over permutations of 
$\nu_1,\cdots, \nu_{2n+1}$ weighted by $(-1)^P$} \, , \nonumber \\
&&
 \ve^{AB} 
= \eps^{\mu_1\cdots \mu_{2n+1} \nu_1\cdots \nu_{2n+1}}\, ,
\een
and,
\be \label{emid}
\MM \equiv (\zeta-\ve) \left\{
(e\zeta - 1)  
\left( 1 + {1\over 2} (1 +\zeta \ve ) (e\zeta - 1) \right)^{-1} \zeta \right\}
\, (\ve +\zeta)
\, .
\ee
Since $Q$ and $Y$ are $(2n+1)$-forms, they can be represented as vectors carrying the index $A$.
The terms in the second line of \refb{efin} have to be interpreted this way, {\it e.g.} we have
\be
Q^T\, \MM\, Q \equiv Q_A \MM^{AB} Q_B \equiv \sum_{\mu_1,\cdots ,\mu_{2n+1}\atop \mu_1<\mu_2<\cdots <\mu_{2n+1}}
\sum_{\nu_1,\cdots ,\nu_{2n+1}\atop \nu_1<\nu_2<\cdots <\nu_{2n+1}} Q_{\mu_1\cdots \mu_{2n+1}}
\MM^{\mu_1\cdots \mu_{2n+1}; \nu_1\cdots \nu_{2n+1}} Q_{\nu_1\cdots \nu_{2n+1}}\, .
\ee
It was shown in \cite{1511.08220} that the equations of motion derived from the action 
\refb{e4.0}, together with the self-duality
constraint \refb{e4con}, are the same as the equations of motion derived from the action 
\refb{e4tot} if we make the
identification:
\be \label{eqfid}
Q = (1+\zeta\ve) F = (F+*F), \quad F={1\over 2} (Q-R)\, ,
\ee
with $R$ defined as in \refb{edefR}.

The action \refb{e4tot} has an obvious gauge symmetry $P\to P+d\, \Xi$ where $\Xi$ is a $(2n-1)$-form gauge transformation
parameter. Typically the original action \refb{e4.0} has another gauge symmetry under which $Y$ 
transforms as a Chern-Simons form. The complete transformation laws take the form
\be 
Y \to Y+d\Lambda, \quad C\to C-\Lambda\, ,
\ee
where $\Lambda$ is a combination of some gauge transformation parameter and other fields, satisfying
\be
\int \Lambda\wedge dY=0\, .
\ee
The new action \refb{e4tot} can be
shown to be invariant under the same gauge symmetry provided we assign the following transformations laws
to $P$ and $Q$:
\be
\delta P = -\Lambda, \quad \delta Q=-(d\Lambda+*d\Lambda)\, .
\ee
The combinations that remain invariant under this gauge transformation are
\be
dP+Y, \qquad Q+Y+*Y\, .
\ee

The new action \refb{e4tot} also has infinitesimal 
general coordinate invariance if we assign appropriate transformation laws
for $P$ and $Q$. The explicit form of these transformation laws can be found in 
\cite{1511.08220}. These can be integrated to generate finite general 
coordinate transformations
that are continuously connected to identity transformation. In section \ref{s5} we shall
also see examples of the invariance of the theory under diffeomorphisms
that are not connected to identity transformation. This will include diffeomorphisms 
of a torus $T^2$, 
generating S-duality symmetry of a theory of chiral 2-forms in six dimensions
compactified on $T^2$.

\sectiono{Canonical formalism} \label{s3}

Decoupling of the free field part from the interacting part of the theory is clear from the
equations of motion \refb{esec} and \refb{eqeq}. 
Indeed for given $Q$, the degrees of freedom of $P$
satisfy the free field equation of motion encoded in the first equation of
\refb{esec}. Furthermore the particular solution for
$P$ that we choose does not affect the interacting field equations \refb{eqeq} for $Q$.
The formulation of the full BV quantized type II string field theory shows that quantum effects
do not affect this decoupling.
Nevertheless one may find it uncomfortable that the decoupling cannot be seen at the
level of the action. For this reason we shall now study the theory in the canonical 
formalism and show that the Hamiltonian splits into a sum of two Hamiltonians 
containing independent degrees of freedom, and one of them is free. The physical
interacting degrees of freedom reside in the other Hamiltonian.

Our starting point will be the action \refb{eact}.
We take the independent components of $P$ to be
$P_{i_1\cdots i_{2n}}$, $P_{0i_1\cdots i_{2n-1}}$ with 
$1\le i_1<i_2<\cdots < i_{2n-1}<i_{2n}$. 
Making use of the self-duality of
$Q$, 
\be
Q^{0i_1\cdots i_{2n}} = \eps^{0i_1\cdots i_{2n} j_1\cdots j_{2n+1}} Q_{j_1\cdots j_{2n+1}}
\equiv \eps^{i_1\cdots i_{2n} j_1\cdots j_{2n+1}} Q_{j_1\cdots j_{2n+1}}\, ,
\ee
we take
the independent components of  $Q$ to be
$Q_{i_1\cdots i_{2n+1}}$ for $1\le i_1<i_2<\cdots i_{2n+1}\le 4n+1$. 
In terms of these variables the Lagrangian associated with the action \refb{eact}
takes the form:
\ben \label{elag1}
L &=& -\, {1\over 2} \sum_{i_1,\cdots, i_{2n}\atop i_1<i_2\cdots < i_{2n}}\int d^{4n+1}x\,  (\p_0 P_{i_1\cdots i_{2n}}
+\p_{[i_1} P_{i_2\cdots i_{2n}]0})^2
\nonumber \\ &&
+\, {1\over 2}\sum_{i_1,\cdots, i_{2n+1}\atop i_1<i_2<\cdots<i_{2n+1}} \int d^{4n+1}x\, 
(\p_{[i_1} P_{i_2\cdots i_{2n+1}]})^2
\nonumber \\ &&
- \sum_{i_1,\cdots i_{2n}\atop i_1<i_2<\cdots<i_{2n}}\sum_{ j_1 \cdots, j_{2n+1}\atop
j_1<j_2<\cdots <j_{2n+1}} \int d^{4n+1}x\, 
\eps^{i_1\cdots i_{2n}j_1\cdots j_{2n+1}}\{\p_0 P_{i_1\cdots i_{2n}}
+\p_{[i_1} P_{i_2\cdots i_{2n}]0}\} Q_{j_1\cdots j_{2n+1}}
\nonumber \\ &&
-\sum_{i_1,\cdots, i_{2n+1}\atop i_1<i_2<\cdots<i_{2n+1}} \int d^{4n+1}x\, 
\p_{[i_1} P_{i_2\cdots i_{2n+1}]} Q^{i_1\cdots i_{2n+1}} +\int d^{4n+1}x\, \LL_I(Q,\Psi)\, ,
\een
where
\ben 
&& \p_{[i_1} P_{i_2\cdots i_{2n+1}]} \equiv \p_{i_1} P_{i_2\cdots i_{2n+1}}+\hbox{ cyclic permutations 
of $i_1,\cdots, i_{2n+1}$ with sign},
\nonumber \\
&& \p_{[i_1} P_{i_2\cdots i_{2n}]0} \equiv \p_{i_1} P_{i_2\cdots i_{2n}0}+\hbox{ cyclic permutations 
of $i_1,\cdots, i_{2n}$ with sign}\, .
\een
Since our main interest will be in the fields $P$ and $Q$, while looking for the
canonical formulation of the theory we shall treat $P$ and $Q$ in the canonical
formulation but continue to treat the other fields in the Lagrangian formulation. 
The
resulting Hamiltonian should actually be called the Routhian even though we shall
refer to it as the Hamiltonian. A slightly different perspective will be that we treat
all fields other than $P$ and $Q$ as  background fields. From either viewpoint, we need
to introduce the conjugate momenta $\Pi_P^{i_1\cdots i_{2n}}$,
$\Pi_P^{0 i_1\cdots i_{2n-1}}$ and $\Pi_Q^{i_1\cdots i_{2n+1}}$ 
for $i_1<i_2<\cdots <i_{2n+1}$. They can be computed from the Lagrangian 
\refb{elag1} via standard procedure, and are given as follows:
\ben \label{emomen}
&&\Pi_P^{i_1\cdots i_{2n}} =-(\p_0 P_{i_1\cdots i_{2n}}
+\p_{[i_1} P_{i_2\cdots i_{2n}]0})
- \sum_{ j_1, \cdots, j_{2n+1}\atop
j_1<j_2<\cdots <j_{2n+1}} \eps^{i_1\cdots i_{2n}j_1\cdots j_{2n+1}}
Q_{j_1\cdots j_{2n+1}}, 
\nonumber \\
&& \Pi_Q^{i_1\cdots i_{2n+1}}=0, \qquad \Pi_P^{0 i_1\cdots i_{2n-1}}=0\, .
\een
The Hamiltonian is given by
\ben \label{eham}
\RR &=&
\sum_{i_1,\cdots, i_{2n}\atop i_1<i_2\cdots < i_{2n}}
 \int d^{4n+1}x\,  \Pi_P^{i_1\cdots i_{2n}} \p_0 P_{i_1\cdots i_{2n}}
 - L
\nonumber \\
&=& -{1\over 2} \sum_{i_1,\cdots, i_{2n}\atop i_1<i_2\cdots < i_{2n}}\int d^{4n+1}x\,  
\left(\Pi_P^{i_1\cdots i_{2n}} +
\sum_{ j_1 \cdots j_{2n+1}\atop
j_1<j_2<\cdots <j_{2n+1}} \eps^{i_1\cdots i_{2n}j_1\cdots j_{2n+1}}
Q_{j_1\cdots j_{2n+1}} \right)^2
\nonumber \\ &&
- \sum_{i_1,\cdots, i_{2n+1}\atop i_1<i_2<\cdots<i_{2n+1}} \int d^{4n+1}x\, 
 \Pi_P^{i_1\cdots i_{2n}} 
 \p_{[i_1} P_{i_2\cdots i_{2n}]0}
 - {1\over 2}\sum_{i_1,\cdots, i_{2n+1}\atop i_1<i_2<\cdots<i_{2n+1}} \int d^{4n+1}x\, 
(\p_{[i_1} P_{i_2\cdots i_{2n+1}]})^2
\nonumber \\ &&
+\sum_{i_1,\cdots, i_{2n+1}\atop i_1<i_2<\cdots<i_{2n+1}} \int d^{4n+1}x\, 
\p_{[i_1} P_{i_2\cdots i_{2n+1}]} Q^{i_1\cdots i_{2n+1}} - \int d^{4n+1}x\, \LL_I(Q,\Psi)\, .
\een
In our analysis below we shall not write down the
range of summations over the repeated 
indices $i_1,i_2,\cdots$ explicitly, but it will be understood that
whenever the indices of an antisymmetric tensor like $Q_{i_1\cdots i_{2n+1}}$ or
$\p_{[i_1}P_{i_2\cdots i_{2n+1}]}$ are summed, the sum runs only
over inequivalent sets like $i_1< i_2\cdots < i_{2n}<i_{2n+1}$.

Equations in the second line of \refb{emomen} are constraint equations. 
Therefore the canonical formulation of the theory requires the use of 
Dirac's procedure for constrained systems. For the sake of brevity we shall describe below
the main steps in this analysis instead of the full details. 

The Poisson 
brackets between the constraints in the second line of \refb{emomen} vanish, but 
the Poisson brackets of these constraints with $\RR$ do not vanish, leading to
secondary constraints. The Poisson bracket between $\Pi_Q^{j_1\cdots j_{2n+1}}$
and $\RR$ yields the constraint:
\be \label{esol1}
\chi^{j_1\cdots j_{2n+1}}=0\, ,
\ee
where
\be \label{esol2}
\chi^{j_1\cdots j_{2n+1}} \equiv
\eps^{i_1\cdots i_{2n}j_1\cdots j_{2n+1}} 
\Pi_P^{i_1\cdots i_{2n}} +
Q_{j_1\cdots j_{2n+1}}
- \p_{[j_1} P_{j_2\cdots j_{2n+1}]} + {\p \LL_I\over \p Q_{j_1\cdots j_{2n+1}}}\,.
\ee
On the other hand, the Poisson bracket of $\Pi_P^{0 i_2\cdots i_{2n}}$ with $\RR$ leads to 
the constraint:
\be \label{esol53}
\p_{i_1} \Pi_P^{i_1\cdots i_{2n}} =0\, .
\ee

If we define
\be\label{edefpipm}
\Pi_{\pm}^{i_1\cdots i_{2n}} \equiv 
{1\over 2} \left(\Pi_{P}^{i_1\cdots i_{2n}}\pm \eps^{i_1\cdots i_{2n} j_1\cdots j_{2n+1}}
\p_{[j_1} P_{j_2\cdots j_{2n+1}]}
\right)\, ,
\ee
then the constraint \refb{esol53} gives,
\be\label{e3.17}
\p_{i_1} \Pi_\pm^{i_1\cdots i_{2n}} =0\, ,
\ee
and the constraint
\refb{esol1}
can be written as
\be\label{e39}
\Pi_-^{i_1\cdots i_{2n}} = -{1\over 2} \eps^{i_1\cdots i_{2n}j_1\cdots j_{2n+1}} \left[
Q_{j_1\cdots j_{2n+1}} + {\p \LL_I\over \p Q_{j_1\cdots j_{2n+1}}}
\right]\, .
\ee
We shall express this as
\be \label{eform1}
Q_{k_1\cdots k_{2n+1}}- f_{k_1\cdots k_{2n+1}} \left(
\Pi_-,\Psi\right)=0\, ,
\ee
for some function $f_{k_1\cdots k_{2n+1}}$, obtained by solving \refb{e39}.  
We now note that the Poisson bracket of the left hand side of \refb{eform1}
and  $\Pi_Q^{i_1\cdots i_{2n+1}}$
does not vanish:
\be \label{esec22}
\{ Q_{k_1\cdots k_{2n+1}}(t,\vec x)- f_{k_1\cdots k_{2n+1}} \left(
\Pi_-(t,\vec x),\Psi(t,\vec x)\right), \Pi_Q^{i_1\cdots i_{2n+1}}(t, \vec y)\}
=\delta^{i_1}_{k_1}\cdots \delta^{i_{2n+1}}_{k_{2n+1}}\, \delta(\vec x-\vec y)
\, .
\ee
Therefore the constraints $\Pi_Q^{i_1\cdots i_{2n+1}}=0$ 
and \refb{eform1} are second class constraints, and
we can eliminate $\Pi_Q^{i_1\cdots i_{2n+1}}$ and $Q_{i_1\cdots i_{2n+1}}$ 
using these constraints.
The Hamiltonian $\RR$ given in \refb{eham} can now be
expressed as:
\ben \label{eform2}
\RR &=& - \int d^{4n+1}x\, \Pi_+^{i_1\cdots i_{2n}}  \Pi_+^{i_1\cdots i_{2n}}  
+  \int d^{4n+1}x\,  \Pi_-^{i_1\cdots i_{2n}}  \Pi_-^{i_1\cdots i_{2n}}  + 
\int d^{4n+1}x\,g(\Pi_-,\Psi)\nonumber \\ &&
- \int d^{4n+1}x\, \Pi_P^{i_1\cdots i_{2n}} 
\p_{[i_1} P_{i_2\cdots i_{2n}]0}\, ,
\een
where
\be \label{edefgfin}
g(\Pi_-,\Psi)=\left[-{1\over 2} \sum_{ j_1 \cdots j_{2n+1}\atop
j_1<j_2<\cdots <j_{2n+1}} 
\left(
{\p \LL_I\over \p Q_{j_1\cdots j_{2n+1}}}\right)^2
- \LL_I(Q,\Psi)\right]\, ,
\ee
evaluated at the value of $Q$ obtained by solving \refb{e39}:
\ben \label{esol2x}
Q_{j_1\cdots j_{2n+1}} + 2\, 
\eps^{i_1\cdots i_{2n}j_1\cdots j_{2n+1}} 
\Pi_-^{i_1\cdots i_{2n}} + {\p \LL_I\over \p Q_{j_1\cdots j_{2n+1}}}=0\,.
\een
The Poisson brackets of \refb{esol53} with $\RR$, $\Pi_Q^{i_1\cdots i_{2n+1}}$ and 
$\Pi_P^{0 i_1\cdots i_{2n-1}}$, as well as with the secondary constraint
\refb{eform1}, vanish. Therefore there are
no new constraints. We can now regard $\Pi_P^{0 i_1\cdots i_{2n-1}}$ and
$\p_{i_1} \Pi_P^{i_1\cdots i_{2n}}$
as first class
constraints, generating gauge transformation. 

Let us now take stock of the independent phase space degrees of freedom.
After elimination of $Q_{i_1\cdots i_{2n+1}}$ and
$\Pi_Q^{i_1\cdots i_{2n+1}}$ using the second class constraints, we are left with
the independent phase space 
degrees of freedom $P_{i_1\cdots i_{2n}}$,
$\Pi_P^{i_1\cdots i_{2n}}$, $P_{0i_1\cdots i_{2n-1}}$ and $\Pi_P^{0i_1\cdots i_{2n-1}}$.
Of these $\Pi_P^{0i_1\cdots i_{2n-1}}$ and $\p_{i_1} \Pi_P^{i_1\cdots i_{2n}}$  
are set to zero by the first class constraints. Furthermore gauge transformations
generated by these first class constraints involve arbitrary shifts of $P_{0i_1\cdots i_{2n-1}}$
and the shift of $P_{i_1\cdots i_{2n}}$ by $\p_{[i_1}\Lambda_{i_2\cdots i_{2n}]}$
for arbitrary $(2n-1)$-form $\Lambda$.
One can use this gauge symmetry to set to zero
the components $P_{0i_1\cdots i_{2n-1}}$ and the components of
$P_{i_1\cdots i_{2n}}$ proportional to
$\p_{[i_1}\Lambda_{i_2\cdots i_{2n}]}$.
The left-over gauge invariant degrees of freedom can be taken to be
$\Pi_\pm^{i_1\cdots i_{2n}}$ defined in 
\refb{edefpipm}, subject to the constraint \refb{e3.17}.

It is easy to see that the Dirac bracket between any two gauge invariant
quantities, after being expressed as functions of $\Pi_\pm$ using the second class constraints,
can be computed using the
Poisson brackets of $\Pi_\pm$. 
This can be 
summarized by writing:
\ben \label{ediracb}
&& \{ \Pi_+^{i_1\cdots i_{2n}}(t,\vec x), \Pi_+^{k_1\cdots k_{2n}}(t, \vec y)\}_{DB}
= {1\over 2} \eps^{i_1\cdots  i_{2n} j_1 k_1\cdots k_{2n}} {\p\over \p x^{j_1}} \delta(\vec x-\vec y) \, ,\nonumber \\
&& \{ \Pi_-^{i_1\cdots i_{2n}}(t,\vec x), \Pi_-^{k_1\cdots k_{2n}}(t, \vec y)\}_{DB}
= -{1\over 2} \eps^{i_1\cdots  i_{2n} j_1 k_1\cdots k_{2n}} {\p\over \p x^{j_1}} \delta(\vec x-\vec y) \, ,\nonumber \\
&& \{ \Pi_+^{i_1\cdots i_{2n+1}}(t,\vec x), \Pi_-^{j_1\cdots j_{2n+1}}(t, \vec y)\}_{DB}=0\, .
\een

The net outcome of this analysis is that 
the total Hamiltonian becomes the sum of a free Hamiltonian $\RR_+$ and an
interacting Hamiltonian $\RR_-$ given by
\be \label{er+r-}
\RR_+=- \int  d^{4n+1}x\,  \Pi_+^{i_1\cdots i_{2n}}  \Pi_+^{i_1\cdots i_{2n}}, \quad
\RR_- =  \int  d^{4n+1}x\, \Pi_-^{i_1\cdots i_{2n}}  \Pi_-^{i_1\cdots i_{2n}}  + 
\int  d^{4n+1}x\, g(\Pi_-,\Psi)\, ,
\ee
with the variables $\Pi_\pm$ subject to the constraints \refb{e3.17} and the Dirac bracket
\refb{ediracb}. 
In arriving at \refb{er+r-} we have set to zero terms proportional to the first class constraints.
$\RR_+$ is negative definite. However this has no
physical relevance since it decouples from the interacting degrees of freedom $\Pi_-$ and $\Psi$.

The equations of motion derived from this Hamiltonian take the form:
\be \label{ehameq}
\p_t \, \Pi_{\pm} = \{ \Pi_\pm, \RR_{\pm}\}_{DB}\, ,
\ee
and
\be \label{erouth}
{\delta \over \delta \Psi} \int \, dt \, \RR_- = 0\, .
\ee
Eq.\refb{erouth} reflects the fact that $\RR_++\RR_-$ actually describes the Routhian in which
the equations of motion for $\Psi$ are obtained from the Euler-Lagrangian equations. 
Eqs.\refb{ehameq}, \refb{erouth}, together with the constraints \refb{e3.17}, \refb{esol2x},
can be shown to be equivalent to the 
original equations of motion \refb{esec}, after we use the relation between $\Pi_\pm$ and the
original variables $P,Q$ given in \refb{emomen}, \refb{edefpipm}.


\sectiono{Chiral scalar in two dimensions and its compactification on $S^1$} \label{s4}

Let us consider the case of a chiral scalar in two dimensions. This system has been analyzed
using various approaches earlier\cite{siegel,jackiw,schwimmer,costa,bernstein,labastida,abdalla,hull,
gates1,gates2,gates3}, 
but our goal will be to examine how the formalism developed
in \S\ref{s3} can be applied to this case.
In this case $P$ is a scalar $\phi$ and $Q$ is a one form $A$ satisfying the condition
\be
A^\mu = \eps^{\mu\nu}A_\nu\, ,
\ee
where $\eps^{01}=-\eps^{10}=1$, $\eps^{00}=\eps^{11}=0$. This gives
\be \label{eareln}
A^0=A_1\equiv A, \quad A_0=-A^0=-A\, .
\ee
We shall take $A=A_1$ as the independent degree of freedom.
The action \refb{eact} takes the form
\be \label{eaction}
S=\int d^2x \left[-{1\over 2} \left\{ (\p_0\phi)^2 - (\p_1\phi)^2  \right\} - (\p_0\phi + \p_1\phi)A
+ \LL_I(A,\Psi)\right]\, ,
\ee
where $\Psi$ represents other degrees of freedom.

We can construct the Hamiltonian of the system following the procedure described in \S\ref{s3}. Eqs.
\refb{emomen}, \refb{edefpipm} and \refb{e39} take the form:
\be
\Pi_\phi = -\p_0\phi - A\, , \quad \Pi_A=0\, ,
\ee
\be \label{edefppm2d}
\Pi_\pm = {1\over 2} \left(\Pi_\phi\pm \p_1\phi\right)\, ,
\ee
and
\be \label{econs2d}
2\, \Pi_- + A + {\p \LL_I\over \p A}=0\, .
\ee
Eq. \refb{eform2} now gives the form of the Hamiltonian to be
\be
\RR = \RR_++\RR_-\, ,
\ee
where
\be \label{edegh2d}
\RR_+= -  \int dx^1 (\Pi_+)^2, \quad
\RR_-= \int dx^1 \left[ 
(\Pi_-)^2 +g\left(\Pi_-,\Psi\right)
\right]\, ,
\ee
and $g(\Pi_-,\Psi)$ is given by \refb{edefgfin}, \refb{esol2x}:
\be \label{egdef2d}
g(\Pi_-, \Psi) = -\left[{1\over 2} \left({\p\LL_I\over \p A}\right)^2 + 
\LL_I\right]\, ,
\ee
evaluated at the solution to  \refb{econs2d}.
Finally, \refb{ediracb} gives the Dirac brackets between $\Pi_\pm$:
\ben
&& \{\Pi_+(x^0, x^1),\Pi_+(x^0, y^1)\}_{DB} = {1\over 2} \, \delta'(x^1-y^1), 
 \quad \{\Pi_+(x^0, x^1),\Pi_-(x^0, y^1)\}_{DB} =0, \nonumber \\ &&
\{\Pi_-(x^0, x^1),\Pi_-(x^0, y^1)\}_{DB} =- {1\over 2} \, \delta'(x^1-y^1) \, .
\een
$\RR_+$ describes a free theory decoupled from the rest of the fields $\Pi_-$ and
$\Psi$, including the
metric. On the other hand,
$\RR_-$ describes an interacting theory of $\Pi_-$ coupled 
to other fields. Therefore the spectrum of
$\RR_-$ can be regarded as the physical spectrum
of the theory. In two space-time dimensions, the constraints \refb{e3.17} do not exist since
$P$ and $\Pi_\pm$ are scalars.

We shall now consider compactification of the theory on a circle of radius
$R$. While putting it on a circle is achieved by taking the 
$x^1$ coordinate to be compact with fixed
period (which we take to be $2\pi$), the introduction of the radius $R$ requires 
coupling the theory to gravity. For simplicity we shall set to zero 
all fields other than $\phi$, $A$ and the metric (which we shall treat as background). We allow
$\phi$ to carry winding charge so that the zero modes of $\Pi_\pm$ given in \refb{edefppm2d} 
can be treated as
independent variables.
According to \refb{e4tot}, \refb{efin},
$\LL_I$ in this case will be given by:
\be
\LL_I={1\over 16} \, Q^T     \MM \, 
 Q  \, ,
\ee
with $\MM$ given by eq.\refb{emid} and $Q_0=-Q_1 = -A$.
The evaluation of $\MM$ is simplified by noting that since
the indices $A,B$ in \S\ref{s2} correspond to fully anti-symmetric combination of $2n+1$ Lorentz
indices, for $n=0$ this just corresponds to a Lorentz vector. If we take the background metric to be
\be
g = \pmatrix{-1 & 0\cr 0 & R^2}\, ,
\ee
then the corresponding symmetric zweibein will be
\be
\hat e = \pmatrix{-1 & 0\cr 0 & R}.
\ee
This gives, from \refb{evdef},
\be
\zeta = \pmatrix{-1 & 0\cr 0 & 1}, \quad
e = \pmatrix{-1 & 0\cr 0 & R}, \quad
E =  \pmatrix{-1 & 0\cr 0 & R^{-1}}, \quad
\ve = \pmatrix{0 & 1\cr -1 & 0}\, .
\ee
Substituting these into \refb{emid} we get
\be
\MM={2}\, {R-1\over R+1}\, \pmatrix{1 & -1\cr -1 & 1}\, .
\ee
Therefore
\be \label{e4.16}
\LL_I(A,R) = {1\over 16}  Q_A  \MM^{AB} \, Q_B  = {1\over 2} A^2 {R-1\over R+1}\, ,
\ee
where we have used \refb{eareln}, and that $Q_1=-Q_0$ is called $A$ here.
Eq.\refb{econs2d} now becomes
\be
2\, \Pi_-+A  +{R-1\over R+1} A=0\, ,
\ee
leading to
\be \label{e4.18}
A=-{R+1\over R} \Pi_-\, .
\ee
\refb{egdef2d}, \refb{e4.16} and \refb{e4.18} give:
\be 
g(\Pi_-, R) = -\left[ {1\over 2} \left({R-1\over R+1}\right)^2 A^2 + {1\over 2} A^2 {R-1\over R+1}\right]
= -{R-1\over R} (\Pi_-)^2\, .
\ee
Substituting this into \refb{edegh2d} we finally get
\be\label{errmfin}
\RR_-= {1\over R}\int dx^1  (\Pi_-)^2 \, .
\ee

Let us also compute the momentum $K$, defined as the Noether charge associated with the translation symmetry
along $x^1$. From \refb{eaction}, this is given by
\be
K = \int dx^1 (\p_0\phi +A) \p_1\phi = -\int dx^1 \Pi_\phi \p_1\phi = - \int dx^1 (\Pi_+)^2 
+  \int dx^1 (\Pi_-)^2\equiv K_+ + K_-\, .
\ee
Comparing this with \refb{errmfin} we see that we have $K_-=R\, \RR_-$ as expected for a chiral boson. The factor
of $R$ on the right hand side is a reflection of the fact that in the presence of the background metric, the physical momentum  is given by $K_-/R$. We also have
$K_+=\RR_+$, showing that the degrees of freedom associated with the decoupled free fields carry the same
chirality as the interacting degrees of freedom.

Quantization of this theory is straightforward. We focus on the interacting degrees of freedom $\Pi_-$.
First we replace the Dirac bracket relations by $-i$ times the commutators:
\be \label{ecomm}
[\Pi_-(x^0, x^1),\Pi_-(x^0, y^1)] =-  {i\over 2} \, \delta'(x^1-y^1) \, .
\ee
Since $x^1$ has period $2\pi$, we can expand $\Pi_-(x^0, x^1)$ as:
\be 
\Pi_-(x^0, x^1) = {1\over 2\sqrt{\pi}} \sum_n \alpha_{n}(t) e^{i n x^1}, \quad t\equiv  x^0, \quad
\alpha_n(t)^\dagger = \alpha_{-n}(t)\, .
\ee
\refb{ecomm} now gives
\be
[\alpha_n(t), \alpha_m(t)] = n \, \delta_{m+n,0}\, .
\ee
Defining
\be 
a_n = {1\over \sqrt n} \alpha_n, \quad a_n^\dagger = {1\over \sqrt n} \alpha_{-n}, \quad n> 0\, ,
\ee
we get
\be
[a_n, a_m^\dagger] = \delta_{m,n}, \quad [a_n, a_m]=0, \quad [a_n^\dagger, a_m^\dagger]=0\, ,
\ee
and
\be \label{echiralH}
\RR_- = {1\over R} \sum_{n=1}^\infty n \, a_n^\dagger a_n + {1\over 2 R} \alpha_0^2\, .
\ee
Positivity of $\RR_-$ and hence of $K_-=R\, \RR_-$ shows that the excitations all carry positive
integer momentum along $x^1$, with energy quantized in units of $1/R$, 
as is expected of a chiral boson. More generally the first term of the
Hamiltonian given in \refb{echiralH} gives the correct non-zero mode
spectrum of chiral boson on a circle, with 
$a_n^\dagger$ describing the creation operator of a quantum of energy $n/R$.

Since the variable $\alpha_0$ commutes with every other operator, it can be regarded as a classical variable.
To find its interpretation we note that $\alpha_0$ is the zero mode of $\Pi_-$, which is related to the original
variables $(A,\phi)$ via:
\be \label{epizero}
\Pi_-= -{R\over R+1} A\, .
\ee
Therefore $\alpha_0$ is the $x^1$ independent 
mode of $A$ up to a normalization. Using the identification
\refb{eqfid} we can express \refb{epizero} as
\be
\Pi_-= -{R\over R+1} (F_1 - F_0)\, ,
\ee
where $F$ is some 1-form field strength. Typically $F$ would satisfy a self-duality constraint with
respect to physical metric (the analog of \refb{e4con} with $Y=0$), which gives
\be
F_0 = - F^0 = - R^{-1} \, F_1\, .
\ee
Therefore we get 
\be 
\Pi_- = - F_1\, .
\ee
Typically $F_1$ will satisfy a quantization condition in a physical theory. This tells us that the $\alpha_0$ is
quantized in integer multiples of a constant. 

Given the spectrum, one can construct the partition function defined as\footnote{The normalization  constats
multiplying $\tau_1$ and $\tau_2$ have been adjusted so that {\it had we been able to calculate this from
Euclidean path integral on a torus}, the torus would be defined via the identification $z\equiv z+2\pi R\equiv 
z+2\pi R(\tau_1+i\tau_2)$ with metric $|dz|^2$.}
$Tr\{e^{-2\pi R \tau_2 \RR_-  + 2\pi i \tau_1 
P_-}\}$ = $Tr\{e^{2\pi i R (\tau_1+ i\tau_2) \RR_- }\}$. As is well known, this partition function is not modular invariant. From the perspective of our
analysis this is not a surprise, since the partition function is not computed via Euclidean path integral
on a torus. Indeed, since one of the variables in our theory is a real self-dual vector field $A_\mu$, it is not
clear how to formulate the theory in the Euclidean space.

\sectiono{Chiral 2-form in six dimensions and its dimensional reduction on $T^2$} \label{s5}

In this section we shall analyze the theory of chiral 2-form in six dimensions with 
self-dual 3-form field strength.\footnote{Various aspects of the 
theories of self-dual tensors 
in six dimensions have been discussed in 
\cite{9611065,9701166,9702008,9709014,9712059,9902171,1104.4040,
1108.4060,1108.5131,1203.4224,1209.3017,1406.5185,1804.05059}.}
We begin by describing the usual formulation of the 
theory where we impose the self-duality constraint
after deriving the equations of motion from the action.
We denote by $B$ a 2-form field and by $\Psi$ all other fields including the metric
$g_{\mu\nu}$, and define
\be \label{e5.-1}
H = dB + \Omega(\Psi), 
\ee
where $\Omega$ is a 3-form constructed from the other fields. 
The action is taken to be of the form:
\be\label{e5.0}
S = S_1 + S_2(\Psi)\, ,
\ee
where
\be \label{e5.1}
S_1 = \int \left[-{1\over 2} H\wedge *_g H 
+ H \wedge \Omega 
\right]\, ,
\ee
and $S_2$ depends on fields other than $B$. After deriving the equations of motion from 
\refb{e5.1} by regarding $B_{\mu\nu}$ as an unconstrained field, 
we impose the self-duality constraint on $H$:
\be\label{e5con}
*_g H=H\, .
\ee
Since $d *_gH = dH =d\Omega$, $*_gd\Omega$ describes the string source to which the 2-form field $B$ 
couples.\footnote{Non-dynamical sources
can be represented as terms in the action \refb{e5fin} that are linear in $P$. For example a string along
the $x^1$ direction, situated at the origin of the transverse coordinates, can be 
represented by adding to the action a term proportional to $\int dx^0 dx^1
P_{01}(x^0, x^1, \vec 0_\perp)$.}  
An explicit example
of such an action can be found in \cite{1608.03919}.

Our goal in this section will be to  follow the general procedure reviewed 
in \S\ref{s2} to construct an equivalent
action that avoids the use of the additional constraint \refb{e5con}, 
and then construct the Hamiltonian following the procedure described
in \S\ref{s3}.
Comparison of \refb{e2.2rep}, \refb{e4.0}, \refb{e4.1}, \refb{e4con} 
and \refb{e5.-1}, \refb{e5.0}, \refb{e5.1}, \refb{e5con} shows that 
we have the identification:
\be
Y=\Omega, \quad C=B, \quad F = dB = H-\Omega\, .
\ee
Therefore using \refb{efin} we can replace
the action \refb{e5.1} and the constraint  \refb{e5con} by a new 
action:\footnote{Action for self-dual 3-forms with similar kinetic term has been 
considered by Neil Lambert\cite{private}.}
\be \label{e5fin}
S'_1 = {1\over 2}  \int d P \wedge * d P
- \int d P \wedge Q  + \int d^6 x\,  \LL_I \, ,
\ee
where
\be \label{e57}
\LL_I \equiv -{1\over 2}  Q^T (\zeta-\ve) \Omega 
+ {1\over 16}   Q^T     \MM \,
 Q  
+ {1\over 4}  Q^{T}   \MM \,
\Omega 
-{1\over 2} \Omega^T 
\zeta \,\Omega
+{1\over 4}  \Omega^T  \MM\,  \Omega \, .  \nonumber 
\ee
Here $P$ is a 2-form field and $Q$ is an independent three form field.
It follows from \refb{eqfid} that 
the degrees of freedom in the new description and the original description are 
related by:
\be
Q = dB + * dB\, .
\ee

We now compactify two of the dimensions on $T^2$ with coordinates 
$x^4$, $x^5$, each of period $1$. Also we shall work with a background metric of the
form
\be \label{ebc1}
\pmatrix{\eta &  \cr & G},
\ee
where $\eta$ denotes four dimensional Minkowski metric and  $G$ is a $2\times 2$ 
symmetric matrix
describing the metric on $T^2$. The corresponding symmetric vielbein is given by:
\be 
\hat e = \pmatrix{\eta &  \cr & \Gamma}\, ,
\ee
where $\Gamma$ is a $2\times 2$ symmetric, positive definite matrix satisfying, 
\be \label{egamG}
\Gamma^2=G\, .
\ee 
In the following we shall use the symbols $\alpha,\beta,\cdots$ for labelling the compact directions 
4 and 5,
$\mu,\nu,\rho,\cdots$ for labelling the non-compact directions 0,1,2 and 3, and $i,j,k,\cdots$ for
labelling the non-compact spatial directions $1,2,3$. 
As before, the indices are raised and lowered by the metric $\delta_{\alpha\beta}$,
$\eta_{\mu\nu}$, $\delta_{ij}$
and their inverse.
$\eps_{\alpha\beta}$, 
$\eps^{\mu\nu\rho\sigma}$ and $\eps^{ijk}$
will denote totally anti-symmetric symbols with $\eps_{45}=1$, $\eps^{0123}=1$ and
$\eps^{123}=1$.
We also define:
\be
I\equiv \pmatrix{1 & 0\cr 0 & 1}, \quad \EE\equiv 
\pmatrix{0 & 1\cr -1 & 0}, \qquad 
A\equiv {1\over 2} (I+\Gamma), \quad B \equiv  {1\over 2}\, \EE \, (\Gamma-I)\, ,
\ee
\be
\wt A \equiv  (A+BA^{-1} B)^{-1}, \qquad \wt B \equiv  - A^{-1} B \wt A\, ,
\ee
\be \label{eend}
\KK \equiv  \Gamma\wt A +\EE\,  \Gamma\wt B -\wt A -\EE\wt B -\Gamma \wt B\, \EE +\EE\Gamma
\wt A \, \EE +\wt B \EE-\EE \wt A \, \EE\, ,
\ee
and
\be
\wh \KK \equiv  -\EE \, \KK = \KK \, \EE\, .
\ee
Straightforward algebra yields the following
results for the various components of the matrix $\MM$ defined in \refb{emid}:
\be
\MM^{\mu\nu\rho;\mu'\nu'\rho'} = 2\, {\left\{\det\Gamma
-1\right\} \over \left\{\det\Gamma
+1\right\}} \,  \eta^{\mu\mu'}\eta^{\nu\nu'}\eta^{\rho\rho'}\, ,
\ee
\be 
\MM^{\mu\nu\rho; \sigma 45} = \MM^{\sigma 45;\mu\nu\rho} =
-2\, {\left\{\det\Gamma
-1\right\} \over \left\{\det\Gamma
+1\right\}} \, \eps^{\mu\nu\rho\sigma}\, ,
\ee
\be
\MM^{\mu 45; \nu 45}= 2\, \eta^{\mu\nu} \, {\left\{\det\Gamma
-1\right\} \over \left\{\det\Gamma
+1\right\}} \, ,
\ee
\be
\MM^{\mu\nu\alpha; \rho\sigma\beta}= \KK_{\alpha\beta}\, \eta^{\mu\rho}\,
\eta^{\nu\sigma} + \wh\KK_{\alpha\beta} \, \eps^{\mu\nu\rho\sigma}\, ,
\ee 
and
\be \label{emzero}
\MM^{\mu\nu\alpha;\mu'\nu'\rho'}=0, \quad \MM^{\mu'\nu'\rho';\mu\nu\alpha}=0, 
\quad\MM^{\mu\nu\alpha;\rho 45}=0, \quad\MM^{\rho 45;\mu\nu\alpha}=0\, .
\ee
Using \refb{egamG}-\refb{eend} one can also show that
\be \label{eimprel}
I + {1\over 2} \KK = 2 \, \left(I + \sqrt{\det G} \, G^{-1}\right)^{-1}, \quad I - {1\over 2} \KK = 
2\, \left(I + \sqrt{\det G} \,
G^{-1}\right)^{-1} \, \sqrt{\det G} \, G^{-1}\, .
\ee

Using these results, and the results of \S\ref{s3}, 
we can write down the expression for the full action and the corresponding
Hamiltonian. However since our main goal will be to test S-duality invariance, we shall
work with a truncated theory in which we take all the fields to be independent of the
coordinates $x^4,x^5$.
Eqs.\refb{emzero}  now
shows that the action given in \refb{e5fin}, \refb{e57} 
can be written as a sum of two terms -- one containing the fields
$P_{45}$, $P_{\mu\nu}$, $Q_{45\mu}$, $Q_{\mu\nu\rho}$ and the sources $\Omega_{45\mu}$ and
$\Omega_{\mu\nu\rho}$, and the other containing the fields $P_{\alpha\mu}$, $Q_{\alpha\mu\nu}$,
and the sources $\Omega_{\alpha\mu\nu}$. Therefore 
it is possible to make a consistent truncation of the theory
by setting\footnote{For this argument we are treating $\Omega$ as an independent field so that it is
possible to set $\Omega_{45\mu}$ and $\Omega_{\mu\nu\rho}$ to zero, keeping
$\Omega_{\alpha\mu\nu}$ non-zero..}
\be \label{etrunc}
P_{45}=0, \quad P_{\mu\nu}=0, \quad Q_{45\mu}=0,  \quad 
Q_{\mu\nu\rho}=0, \quad \Omega_{\mu\nu\rho}=0, \quad \Omega_{45\mu}=0\, .
\ee
These fields describe the dynamics of S-duality invariant 
scalar fields and their sources in four dimensions and
are not needed to understand the dynamics of the gauge fields that is of interest to us.
The left-over independent fields / sources are:
\be\label{elist1}
P_{(\alpha)\mu} \equiv -P_{\alpha\mu}, \quad Q_{(\alpha)\mu\nu}
\equiv Q_{\alpha\mu\nu}, \quad
\Omega_{(\alpha)\mu\nu} \equiv \Omega_{\alpha\mu\nu}\, , 
\quad 
\hbox{for $\alpha=4,5$}\, .
\ee
Self-duality of $Q$ in six dimensions implies that in four dimensions:
\be \label{eqdual}
Q_{(\alpha)} = \eps_{\alpha\beta} \star \, Q_{(\beta)}\, ,
\ee
where $\star$ denotes Hodge dual in 3+1 dimensions with respect to the four dimensional
Minkowski metric:
\be
\star A_{\mu\nu} = \eps_{\mu\nu}^{~~\rho\sigma} A_{\rho\sigma}\, .
\ee 
From the perspective of the 3+1 dimensional theory, $\eps_{\alpha\beta} \star d\Omega_{(\beta)}$ represents
the current density of string charge wrapped along $\alpha$ direction.

Using the truncation \refb{etrunc}, and the relation
\be
(dP_{(\alpha)})_{\mu\nu} = (dP)_{\alpha\mu\nu}\, ,
\ee
the action $S_1'$ given in \refb{e5fin} takes the following form
in terms of the fields listed in \refb{elist1}:
\be \label{eredact}
S_1' = \int d^4 x \bigg[{1\over 2} 
(dP_{(\alpha)})_{\mu\nu} (dP_{(\alpha)})^{\mu\nu} 
- (dP_{(\alpha)})_{\mu\nu} (Q_{(\alpha)})^{\mu\nu}  +\LL_I\bigg]\, ,
\ee
with
\ben \label{eliform}
\LL_I &=&
 {1\over 8}  \, \KK_{\alpha\beta}\,  Q_{(\alpha)\mu\nu}    Q_{(\beta)}^{~~\mu\nu}    
- Q_{(\alpha)\mu\nu} \Omega_{(\alpha)}^{~~\mu\nu} 
+  {1\over 2} \, \KK_{\alpha\beta}\, Q_{(\alpha)\mu\nu} 
\Omega_{(\beta)}^{~~\mu\nu} 
-{1\over 2} \Omega_{\alpha\mu\nu} \Omega_{\alpha}^{~\mu\nu}
+ {1\over 4} \Omega_{(\alpha)\mu\nu} \KK_{\alpha\beta} \Omega_{(\beta)}^{~\mu\nu}
\nonumber \\ &&
+{1\over 4} \Omega_{(\alpha)\mu\nu} (\wh \KK)^{\alpha\beta}
\eps^{\mu\nu\rho\sigma} \Omega_{(\beta)\rho\sigma}
 \bigg] \nonumber \\
&=& {1\over 4} \KK_{\alpha\beta}\,  Q_{(\alpha)ij}    Q_{(\beta)}^{~~ij} 
-  Q_{(\alpha)ij} \left(\Omega_{(\alpha)}^{~~ij} - \eps_{\alpha\beta} 
\star\Omega_{(\beta)}^{~~ij}\right)
+  {1\over 2} \, \KK_{\alpha\beta}\, Q_{(\alpha)ij} 
\left(\Omega_{(\beta)}^{~~ij} + \eps_{\beta\gamma} 
\star\Omega_{(\gamma)}^{~~ij}\right)\nonumber \\ &&
-{1\over 2} \left(\Omega_{(\alpha) ij} \Omega_{(\alpha)}^{~ij}
- \star\Omega_{(\alpha) ij} \star\Omega_{(\alpha)}^{~ij}\right)
+ {1\over 4} \Omega_{(\alpha) ij} \, \KK_{\alpha\beta} \,
\Omega_{(\beta)}^{~~ij} 
- {1\over 4} \star\Omega_{(\alpha) ij} \, \KK_{\alpha\beta} \, 
\star\Omega_{(\beta)}^{~~ij} 
\nonumber \\ && 
+ {1\over 2} \Omega_{(\alpha) ij} \, \KK_{\alpha\beta} \, \eps_{\beta\gamma} \,
\star\Omega_{(\gamma)}^{~~ij} 
\, .
\een
In \refb{eredact}, \refb{eliform} and in all subsequent equations, sum over pairs of indices
that belong to an antisymmetric tensor, {\it e.g.} $\mu,\nu$ in \refb{eredact} and the middle
expression of \refb{eliform}, and $i,j$ in the last expression in \refb{eliform}, run over only
inequivalent combinations, i.e.\ $\mu<\nu$ or $i<j$.
In arriving at the last expression in \refb{eliform},
we have used \refb{eqdual} to express $\LL_I$ in terms of the independent
components of $Q_{(\alpha)\mu\nu}$ which we take to be $Q_{(\alpha)ij}$.  $\star
\Omega_{(\alpha)ij}$ in the last expression in \refb{eliform} is given by:
\be \label{e529}
\star \Omega_{(\alpha)ij} = \star \Omega_{(\alpha)}^{~ij} =\eps^{ij0k} \Omega_{(\alpha)0k}
=\eps^{ijk} \Omega_{(\alpha)0k}\, .
\ee

We can now construct the Hamiltonian associated with this system by straightforward application
of the results of \S\ref{s3}. We shall ignore the effect of the additive term $S_2(\Psi)$ in
\refb{e5.0}, and focus on $\RR_-$ since that is the part that describes the
interacting part of the theory.  After defining $\Pi_-^{(\alpha)k}\equiv\Pi_-^{\alpha k}$, 
the constraint \refb{esol2x} takes the form:
\be
Q_{(\alpha) ij} = -2\, \eps_{\alpha\beta} \, \eps_{k ij} \, \Pi_-^{(\beta) k} - {1\over 2} \, \KK_{\alpha\beta}
\, Q_{(\beta)ij}
+\Omega_{(\alpha)ij} - \eps_{\alpha\beta} \, \star \Omega_{(\beta)ij} - {1\over 2}\, \KK_{\alpha\beta} \left( \Omega_{(\beta)ij} + \eps_{\beta\gamma} \star \Omega_{(\gamma)ij}\right)\, .
\ee
This can be used to express $Q_{(\alpha) ij}$ in terms of $\Pi_-^{(\gamma) k}$:
\be \label{esolqij}
Q_{(\alpha) ij} 
= - 2\left(1 +{1\over 2} \KK\right)^{-1}_{\alpha\beta} \, \eps_{\beta\gamma} \,\eps_{k ij} 
\Pi_-^{(\gamma) k} + \left\{ \left(1+{1\over 2}\KK\right)^{-1} \left(1-{1\over 2}\KK\right)\right\}_{\alpha\beta}
\Omega_{(\beta)ij}- \eps_{\alpha\beta} \star \Omega_{(\beta)ij}\, .
\ee
$\RR_-$ given in \refb{er+r-}, 
\refb{edefgfin}, after using \refb{eimprel}, \refb{etrunc}, \refb{elist1}, \refb{eliform},
\refb{e529} and
\refb{esolqij}, takes the form:
\ben \label{es5hamfin}
\RR_-&=&\int d^3 x \, \Bigg[(\det G)^{-1/2} \, \left\{\Pi_{-}^{(\alpha)i} +\eps_{\alpha\gamma} \Omega_{(\gamma)k\ell}
\eps_{ik\ell}
\right\} G_{\alpha\beta} \left\{\Pi_{-}^{(\beta)i} +\eps_{\beta\delta} \Omega_{(\delta)k'\ell'} \eps_{ik'\ell'}\right\}
\nonumber \\
&& + 2 \, \Pi_{-}^{(\alpha)k}\, 
\Omega_{(\alpha)0k} - \eps_{\alpha\beta} \, \eps_{ijk}\, \Omega_{(\alpha)ij} \Omega_{(\beta)0k}
\Bigg]\, .
\een
On the other hand, the Dirac bracket relation \refb{ediracb} gives:
\be \label{ediracb6d}
\left\{ \Pi_{-}^{(\alpha)i}(t, \vec x) , \Pi_{-}^{(\beta)j}(t, \vec y)\right\}_{DB} = {1\over 2} \, \eps_{\alpha\beta} \eps^{ijk} \, \p_k \delta^{(3)}(\vec x
-\vec y)\, ,
\ee
and the constraint \refb{e3.17} gives:
\be\label{econ6d}
\p_k \, \Pi_-^{(\alpha)k}=0\, .
\ee

We now note that $\RR_-$ is invariant under the S-duality transformation:
\ben \label{eduality}
&& \hskip -.3in G \to W^T \, G\, W, \quad \Omega_{(\alpha)\mu\nu}\to 
(W^T)_{\alpha\beta}  \Omega_{(\alpha)\mu\nu}, \quad 
\Pi_{-}^{(\alpha)i}\to \left(W^{-1}\right)_{\alpha\beta} \, \Pi_{-}^{(\beta)i}, \nonumber \\
&& \hskip 2in \hbox{for} \quad W\in SL(2,\ZZZ)\, .
\een
The Dirac bracket relation \refb{ediracb6d} is also invariant under this transformation since
\be 
W_{\alpha\beta} \eps_{\beta\beta'} W_{\alpha'\beta'} =\eps_{\alpha\alpha'}\, .
\ee

Returning to the action \refb{eredact}, we see that the S-duality transformation, corresponding
to the choice 
\be
W = W_S = \pmatrix{0 & 1\cr -1 & 0}\, ,
\ee
is a symmetry of the action and the self-duality constraint \refb{eqdual} under the transformation
\be
P_{(\alpha)\mu} \to (W_S^T)_{\alpha\beta} P_{(\beta)\mu}, \quad 
\Omega_{(\alpha)\mu\nu}\to 
(W_S^T)_{\alpha\beta}  \Omega_{(\beta)\mu\nu}\, , \quad Q_{(\alpha)\mu\nu} \to 
(W_S^{T})_{\alpha\beta} Q_{(\beta)\mu\nu}, \quad \Gamma\to W_S^T \, \Gamma \, W_S\, .
\ee 
This is a consequence of the fact that the original action is manifestly invariant under
Lorentz transformation, which includes 90$^\circ$ rotation in the 4-5 plane generating
the S transformation. In contrast,
the T transformation generated by
\be
W_T=\pmatrix{1 & 1\cr 0 & 1}\, ,
\ee 
is only a symmetry of the equations of motion (and also of the Hamiltonian) since
it corresponds to a diffeomorphism not connected to the identity. The original action is
invariant under infinitesimal diffeomorphisms which only generate diffeomorphisms connected
to the identity.

\bigskip

{\bf Acknowledgement:} 
I would like to thank the participants at the conference on `String and M-theory, The New Geometry of 21st Century', held at IMS, National University of Singapore, for lively discussions during the
course of this work. 
I would also like to thank the Institute of Mathematical Sciences at the
National University of Singapore for hospitality during this conference.
I would like to thank Neil Lambert for his comments 
on an earlier version of the manuscript and for sharing his unpublished notes on
possible approaches to M5-brane action.
This work was
supported in part by the J. C. Bose fellowship of 
the Department of Science and Technology, India and also by the Infosys Chair Professorship.

\end{document}